\pdfoutput=1
\RequirePackage{ifpdf}
\ifpdf 
\documentclass[pdftex]{sigma}
\else
\documentclass{sigma}
\fi

\numberwithin{equation}{section}

\newtheorem{Theorem}{Theorem}[section]
\newtheorem{Conjecture}[Theorem]{Conjecture}
 { \theoremstyle{definition}
\newtheorem{Definition}[Theorem]{Definition}
}

\def \be {\begin{equation}}
	\def \ee {\end{equation}}
\def \ba {\begin{eqnarray}}
	\def \ea {\end{eqnarray}}

\def \be {\begin{equation}}
	\def \en {\end{equation}}
\def \bes {\begin{eqnarray}}
	\def \ens {\end{eqnarray}}

\def \s {\textsf{s}}
\def \t {\textsf{t}}
\def \p {x}
\def \x {z}
\def \V {\textsf{V}}

\def \GL {{\rm GL}}

\begin{document}
\allowdisplaybreaks

\newcommand{\arXivNumber}{2003.07958}

\renewcommand{\PaperNumber}{078}

\FirstPageHeading

\ShortArticleName{Minimal Kinematics: An All $k$ and $n$ Peek into ${\rm Trop}^+{\rm G}(k,n)$}

\ArticleName{Minimal Kinematics: \\ An All $\boldsymbol{k}$ and $\boldsymbol{n}$ Peek into $\boldsymbol{{\rm Trop}^+{\rm G}(k,n)}$}

\Author{Freddy CACHAZO~$^{\rm a}$ and Nick EARLY~$^{\rm b}$}

\AuthorNameForHeading{F.~Cachazo and N.~Early}

\Address{$^{\rm a)}$~Perimeter Institute for Theoretical Physics,\\
\hphantom{$^{\rm b)}$}~31~Caroline~Str., Waterloo, Ontario N2L 2Y5, Canada}
\EmailD{\href{mailto:fcachazo@perimeterinstitute.ca}{fcachazo@perimeterinstitute.ca}}

\Address{$^{\rm b)}$~The Institute for Advanced Study, Princeton, NJ, USA}
\EmailD{\href{mailto:earlnick@gmail.com}{earlnick@gmail.com}}

\ArticleDates{Received December 14, 2020, in final form August 08, 2021; Published online August 25, 2021}\vspace{-1mm}

\Abstract{In this note we present a formula for the Cachazo--Early--Guevara--Mizera (CEGM) generalized biadjoint amplitudes for all $k$ and $n$ on what we call the {\it minimal kinematics}. We prove that on the minimal kinematics, the scattering equations on the configuration space of $n$ points on $\mathbb{CP}^{k-1}$ has a unique solution, and that this solution is in the image of a Veronese embedding. The minimal kinematics is an all $k$ generalization of the one recently introduced by Early for $k=2$ and uses a choice of cyclic ordering. We conjecture an explicit formula for $m_n^{(k)}(\mathbb{I},\mathbb{I})$ which we have checked analytically through $n=10$ for all $k$. The answer is a simple rational function which has only simple poles; the poles have the combinatorial structure of the circulant graph ${\rm C}_n^{(1,2,\dots, k-2)}$. Generalized biadjoint amplitudes can also be evaluated using the positive tropical Grassmannian ${\rm Tr}^+{\rm G}(k,n)$ in terms of generalized planar Feynman diagrams. We find perfect agreement between both definitions for all cases where the latter is known in the literature. In~particular, this gives the first strong consistency check on the $90\,608$ planar arrays for ${\rm Tr}^+{\rm G}(4,8)$ recently computed by Cachazo, Guevara, Umbert and Zhang. We also introduce another class of special kinematics called {\it planar-basis kinematics} which generalizes the one introduced by Cachazo, He and Yuan for $k=2$ and uses the planar basis recently introduced by Early for all $k$. Based on numerical computations through $n=8$ for all $k$, we conjecture that on the planar-basis kinematics $m_n^{(k)}(\mathbb{I},\mathbb{I})$ evaluates to the multidimensional Catalan numbers, suggesting the possibility of novel combinatorial interpretations. For $k=2$ these are the standard Catalan numbers.}


\Classification{14M15; 05E99; 14T99}

\vspace{-2mm}
\section{Introduction}

The most basic theory which admits a simple Cachazo--He--Yuan formulation (CHY) is the biadjoint scalar theory. This is a theory of a massless scalar field in the adjoint representation of the flavor group $U(N)\times U(M)$ and with only cubic interactions. Color decomposition in both~$U(N)$ and~$U(M)$ leads to $n$-point partial amplitudes that depend on two orderings, $m_n(\alpha, \beta)$. In~this work we are only concerned with the canonical planar ordering $\mathbb{I}:=(1,2,\dots, n)$ and its corresponding partial amplitudes $m_n(\mathbb{I}, \mathbb{I})$.

The CHY formulation of $m_n(\mathbb{I}, \mathbb{I})$ is an integral over the configuration space of $n$ points on~$\mathbb{CP}^{1}$ localized to points satisfying the scattering equations~\cite{Cachazo:2013gna,Cachazo:2013hca,Cachazo:2013iea,Fairlie:2008dg,Fairlie:1972zz}.

Recently, Cachazo, Early, Guevara, and Mizera (CEGM) introduced a generalization of the CHY formulation that uses the configuration space of $n$ points on $\mathbb{CP}^{k-1}$~\cite{Cachazo:2019ngv,Cachazo:2019apa,Cachazo:2019ble}. This also led to generalized biadjoint amplitudes $m_n^{(k)}(\mathbb{I}, \mathbb{I})$. Also noted by CEGM, the beautiful connection between $k=2$ Feynman diagrams entering in the expansion of $m_n^{(2)}(\mathbb{I}, \mathbb{I})$ and the tropical Grassmannian ${\rm Trop}\,{\rm G}(2,n)$ naturally extends to ${\rm Trop}\,{\rm G}(k,n)$. Moreover, if one is only interested in~$m_n^{(k)}(\mathbb{I}, \mathbb{I})$ then it is enough to consider the positive part, ${\rm Trop}^+{\rm G}(k,n)$~\cite{SWTrop}, as~explicitly pointed out in~\cite{Drummond:2019qjk}.

An expansion of $m_n^{(k)}(\mathbb{I}, \mathbb{I})$ in terms of generalized planar Feynman diagrams was introduced by Borges and one of the authors for $k=3$~\cite{Borges:2019csl}, building on the beautiful work of Herrmann, Jensen, Joswig and Sturmfels~\cite{HJJS}, and later extended to arrays of Feynman diagrams to all $k$ by Cachazo, Guevara, Umbert and Zhang (CGUZ)~\cite{Cachazo:2019xjx}.

While the definition of $m_n^{(k)}(\mathbb{I}, \mathbb{I})$ either as a CHY integral or as a sum over planar arrays of~Feynman diagrams is well-understood, its explicit evaluation becomes forbiddingly complicated even for modest values of $k$ and $n$.

In this note we present the first all $k$ and $n$ result obtained by evaluating the CEMG biadjoint amplitudes on what we call the {\it minimal kinematics}. The minimal kinematics is an all $k$ generalization of the one introduced by the second author in~\cite{Early:2018zuw}. The explicit answer is a very compact formula with the combinatorial structure of a circulant graph. Here we see the power of the CHY formula in action as it re-sums large numbers of (generalized) Feynman diagrams into a single compact rational function. We compare our results to the explicit evaluation of the CEMG amplitudes obtained by summing over (generalized) Feynman diagrams and find perfect agreement in the cases that are known in the literature. The most impressive comparisons are those for ${\rm Trop}^+\,{\rm G}(3,8)$ and ${\rm Trop}^+\,{\rm G}(4,8)$, where $13\, 612$ collections and $90\, 608$ matrices all give non-vanishing contributions and their sums collapse to the compact result. This is a very strong consistency check on the ${\rm Trop}^+\,{\rm G}(4,8)$ CGUZ results. The ${\rm Trop}^+\,{\rm G}(3,8)$ CGUZ collections were already checked to reproduce the results obtained by Drummond, Foster, G\"urdo\u{g}an, and Kalousios using cluster algebras in~\cite{Drummond:2019qjk}.

In 2013, CHY noticed that the kinematic invariants of all possible planar poles in a $k=2$ biadjoint amplitude form a basis of the corresponding kinematic space~\cite{Cachazo:2013iea}. Using this fact CHY set all planar kinematic invariants to unity so that each planar Feynman diagram contributes exactly $1$ to the amplitude leading to the result that $m_n^{(2)}(\mathbb{I},\mathbb{I}) = C_{n-2}$ with $C_m$ the $m^{\rm th}$ Catalan number. In~this work we also present a generalization of this {\it planar-basis} kinematics to all $k$ and $n$ using the recently introduced planar basis by the second author in~\cite{Early:2019eun}. We evaluate the CEGM biadjoint amplitude for $k=3$ and $n=5,\,6,\,7,\,8$ and find~$5,\; 42,\; 462,\; 6\, 006$ respectively. These numbers are the first three-dimensional Catalan numbers. We also evaluate $k=4$ and $n=6,\,7,\,8$ and find $14,\; 42,\; 24\, 024$ which are the first four-dimensional Catalan numbers.\footnote{For the general sequence of multi-dimensional Catalan numbers, see~\cite[OEIS~A060854]{oeis}.} This hints that the pattern continues to all values of $k$ and $n$ in which case the amplitude would evaluate to the $k$-dimensional Catalan numbers.

This paper is organized as follows: In Section~\ref{sec:Main results} we summarize the main results of this work. In~Section~\ref{sec:biadjoint amplitudes} we review the construction of the CEGM biadjoint amplitudes. In~Section~\ref{sec:uniqueness} we show that on minimal kinematics there is a single solution to the scattering equations on~$X(k,n)$. In~Section~\ref{sec:reduced determinant} we evaluate the reduced determinant. In~Section~\ref{sec: comparison to tropGkn} we compare the results to those obtained by evaluating planar arrays of Feynman diagrams which are the analog of Feynman diagrams for higher $k$ and correspond to facets of ${\rm Trop}^+\, {\rm G}(k,n)$. In~Section~\ref{sec: induced kinematics} we introduce the notion of next-to-minimal kinematics. In~Section~\ref{sec:PK and Ckn} we introduce and start the study of planar-basis kinematics and its connection to multidimensional Catalan numbers. In~Section~\ref{sec: MK planar basis} we conjecture the expression for the evaluation of the planar basis elements on the minimal kinematics, and in Appendix~\ref{sec:numerical} we conclude with discussions which include some future directions. Appendix~\ref{sec: planar collections 38} contains the evaluation of the planar basis on minimal kinematics while Appendix~\ref{sec: planar matrices 48} contains the evaluation of the $(3,8)$ and $(4,8)$ amplitudes on planar kinematics using their definition as arrays of Feynman diagrams.

\section{Main results}\label{sec:Main results}

The $(k,n)$ space of kinematic invariants is defined in terms of a completely symmetric rank $k$ tensor $\s_{a_1,a_2,\dots, a_k}$ satisfying a $k$-masslessness condition $\s_{b,b,a_3,\dots, a_k}=0$ and $k$-momentum conservation~\cite{Cachazo:2019ngv}
\begin{gather}\label{cond}
\sum_{\substack{a_2,a_3,\dots, a_k=1}}^n\s_{a_1a_2\cdots a_{k}} = 0 \qquad \forall a_1.
\end{gather}
Given an ordering, say $\mathbb{I}=(1,2,\dots, n)$, {\it the minimal kinematics} sets to zero all kinematic invariants except for $2(n-1)$ of them which take values
\begin{gather}
\s_{1,2,\dots, k}=\p_1-\p_k,\
\s_{2,3,\dots, k+1}=\p_2-\p_{k+1},\
\dots,\
\s_{n-1,1,\dots, k-1}=\p_{n-1}-\p_{k-1}, \nonumber
\\
\s_{1,2,\dots, k-1,n} = \p_{k}-\p_{n-1},\
\s_{2,3,\dots, k,n}=\p_{k+1}-\p_1,\
\dots,\
\s_{n-1,1,\dots, k-2,n} =\p_{k-1}-\p_{n-2}.\label{eq:MK}
\end{gather}
The ellipses represent terms obtained by applying a cyclic transformation on the labels in the set $\{1,2,\dots, n-1\}$. In~other words, this kinematics is cyclic with respect to the first $n-1$ labels and hence label $n$ plays a special role.

This kinematics is a generalization of the $k=2$ version introduced by the second author in~\cite{Early:2018zuw},\footnote{In~\cite{Early:2018zuw}, particle $n$ was special because it was made massive while the rest were kept massless.} where continuous (respectively discrete) Laplace transforms were used to compute the simplified expression for $m^{(2)}(\mathbb{I}_n,\mathbb{I}_n)$ when $\alpha'\rightarrow 0$ (respectively, $\alpha'>0$).

As an illustration of the definition, let us write the $k=2$ version more explicitly,
\begin{gather*}
\s_{1,2}=\p_1-\p_2,\quad \s_{2,3}=\p_2-\p_{3},\quad \dots, \quad \s_{n-2,n-1}=\p_{n-2}-\p_{n-1},\quad \s_{n-1,1}=\p_{n-1}-\p_1,
\\
\s_{1,n} =\p_{2}-\p_{n-1},\quad \s_{2,n}=\p_{3}-\p_{1},\quad \dots,\quad \s_{n-2,n}=\p_{n-1}-\p_{n-3},\quad \s_{n-1,n}=\p_1-\p_{n-2}.
\end{gather*}

The reason this is called minimal kinematics is that for any $k$ and $n$ the scattering equations, which are the conditions for finding the critical points of the potential function
\begin{gather*}
{\cal S}_k := \sum_{1\leq a_1<a_2<\cdots <a_k\leq n} \s_{a_1a_2\dots a_k}\log |a_1,a_2,\dots, a_k|,
\end{gather*}
possess a single solution on the minimal kinematics as argued in Section~\ref{sec:uniqueness}. ${\cal S}_k$ serves as a~Morse function on the configuration space of $n$ points in $\mathbb{CP}^{k-1}$. Here we use the standard notation $|a_1,a_2,\dots, a_k|$ for the determinant of the matrix made from the homogeneous coordinates of~points $\{a_1,a_2,\dots, a_k\}$.

In general it is not yet known how many solutions the scattering equations possess. Some known facts are the following: For $k=2$ there are $(n-3)!$ solutions~\cite{Cachazo:2013gna} while for $(k,n)=(3,6)$, $(3,7)$, $(3,8)$ there are $26$, $1\, 272$ and $188\, 112$ solutions respectively~\cite{Cachazo:2019ngv,Cachazo:2019apa,Cachazo:2019ble}.

The minimal kinematics is reminiscent of the dual coordinate space used for the definition of~momentum twistors~\cite{Hodges:2009hk} (for a review see Section~5 of~\cite{Elvang:2013cua}). Here the dual coordinate space is one-dimensional with points $\p_1,\p_2,\dots, \p_{n-1}$ in it. It is convenient to think of this space as a $\mathbb{CP}^1$ and to introduce an extra point denoted $\p_n$. The inhomogenous coordinates can be arranged in a $2\times n$ matrix
\begin{gather*}
\begin{pmatrix}
1 & 1 & \cdots & 1 & 0
\\
\p_1 & \p_2 & \cdots & \p_{n-1} & 1
\end{pmatrix}\!.
\end{gather*}
Note that the auxiliary $n^{\rm th}$ point has been located at infinity.

On the minimal kinematics the single solution is given in terms of a Veronese map from the auxiliary one-dimensional dual kinematic space onto the configuration space of $n$ points on~$\mathbb{CP}^{k-1}$. More explicitly, in a convenient gauge fixing, the solution is then given by
\begin{gather}\label{single}
M_n^{(k)}:=
\begin{pmatrix}
	1 & 1 & \cdots & 1 & 1 & 0 \\
	\p_1 & \p_2 & \cdots & \p_{n-2} & \p_{n-1} & 0 \\[1ex]
	\p_1^2 & \p_2^2 & \cdots & \p_{n-2}^2 & \p_{n-1}^2 & 0 \\
	\vdots & \vdots & \ddots & \vdots & \vdots & \vdots \\
	\p_1^{k-1} & \p_2^{k-1} & \cdots & \p_{n-2}^{k-1} & \p_{n-1}^{k-1} & 1
\end{pmatrix}\!.
\end{gather}

In order to express the result of the explicit computation of the CEGM biadjoint amplitude for any $k$ and $n$ on the minimal kinematics it is convenient to introduce some notation.

A {\it circulant} graph, $C_m^{(l_1,l_2,\dots, l_r)}$, is a graph on $m$ labelled vertices $\{1,2,\dots, m\}$ with edges connecting the $i^{\rm th}$ and $j^{\rm th}$ vertices if and only if $|i-j|\in \{l_1,l_2,\dots, l_r,m-l_1,m-l_2,\dots, m-l_r\}$.\vspace{1ex}

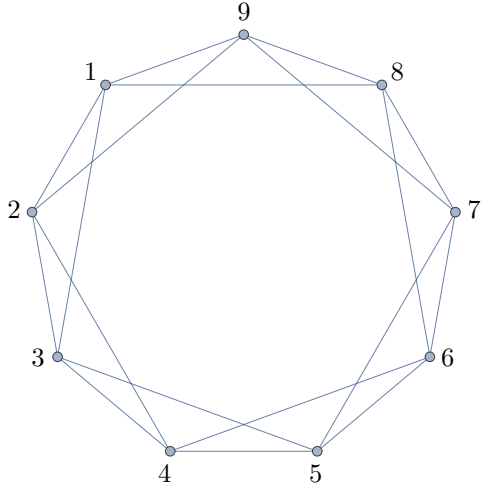
\begin{figure}[ht!]
\centering
\includegraphics{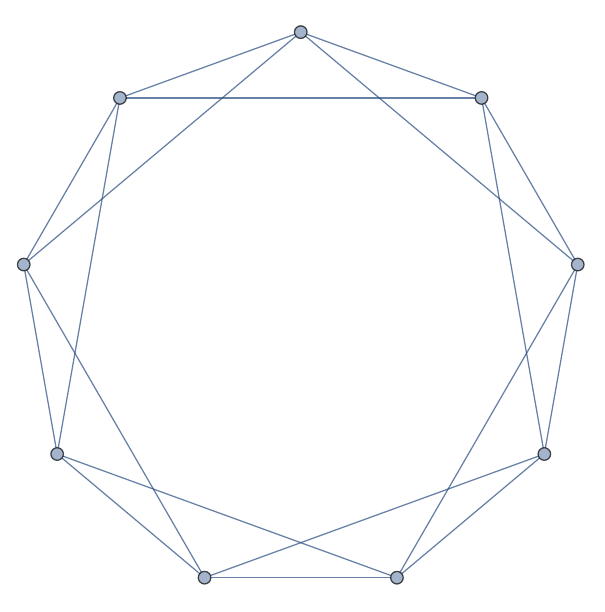}
\put(-143,143){\makebox(0,0)[lb]{\small$1$}}
\put(-172,91){\makebox(0,0)[lb]{\small$2$}}
\put(-163,35){\makebox(0,0)[lb]{\small$3$}}
\put(-115,-9){\makebox(0,0)[lb]{\small$4$}}
\put(-58,-9){\makebox(0,0)[lb]{\small$5$}}
\put(-8,35){\makebox(0,0)[lb]{\small$6$}}
\put(2,91){\makebox(0,0)[lb]{\small$7$}}
\put(-27,143){\makebox(0,0)[lb]{\small$8$}}
\put(-85,166){\makebox(0,0)[lb]{\small$9$}}
\caption{Circulant graph $C_9^{(1,2)}$: This is the graph on $9$ vertices with an edge joining the $i^{\rm th}$ and $j^{\rm th}$ vertices whenever $|i-j|\in \{ 1,2,9-1,9-2\}=\{1,2,8,7\}$.}
\end{figure}

The evaluation of a circulant graph $C_m^{(l_1,l_2,\dots, l_r)}$ on the minimal kinematics is simply a polynomial in $\p_i$'s given by
\begin{gather*}
\big\|C_m^{(l_1,l_2,\dots, l_r)}\big\| := \prod_{e\in E(C)} (\p_{e_a}-\p_{e_b}).
\end{gather*}
The product is over the edge set of $C_m^{(l_1,l_2,\dots, l_r)}$ while $e_a$ and $e_b$ denote the endpoint vertices of the edge $e$. Since $e_a,e_b\in \{ 1,2,\dots, m\}$ we take $e_a<e_b$.

The main result of this work is an explicit formula for the CEGM biadjoint amplitude eva\-lu\-ated on the minimal kinematics:
\begin{gather}\label{formi}
m_n^{(k)}(\mathbb{I},\mathbb{I}) = \frac{\big(R_n^{(k)}\big)^2}{\big\|C_{n-1}^{(1,2,\dots, k-1)}\big\|}
\end{gather}
with
\begin{gather}\label{ratioPT}
R_n^{(k)}:= \frac{|n-(k-1),\dots,n-1,1||n-(k-2),\dots, 1,2|\cdots|n-1,1,2,\dots, k-1|}{|n-(k-1),\dots,n-1,n||n-(k-2),\dots, n,1|\cdots |n,1,\dots, k-1|}
\end{gather}
and $|a_1,a_2,\dots, a_k|$ the $k \times k$ minor of $M^{(k)}_n$, defined in \eqref{single}, formed by the $a_i^{\rm th}$ columns.

The reader familiar with gluon scattering amplitudes can recognize $R_n^{(k)}$ as a generalization of an inverse soft factor. More explicitly,
\begin{gather*}
R^{(2)}_n = \frac{|n-1,1|}{|n-1,n||n,1|}.
\end{gather*}
As an illustration of \eqref{formi}, the $k=3$ case evaluates to
\begin{gather*}
m_n^{(3)}(\mathbb{I},\mathbb{I}) = \frac{(\p_{n-1}-\p_2)^2(\p_{n-1}-\p_{1})^2(\p_{1}-\p_{n-2})^2}{\prod_{i=1}^{n-1}((\p_{i}-\p_{i+1})(\p_{i}-\p_{i+2}))},
\end{gather*}
where the labels in the denominator are defined cyclically in $\{1,2,\dots, n-1\}$.

\section{CEGM biadjoint amplitudes}\label{sec:biadjoint amplitudes}

In this section we review the definition of the CEGM biadjoint amplitude and all the ingredients that have to be computed in order to evaluate the amplitude on the minimal kinematics.

The first ingredient is the $\mathbb{CP}^{k-1}$ scattering equations, i.e., the conditions for finding the critical points of ${\cal S}_k$
\begin{gather*}
\frac{\partial {\cal S}_k}{\partial \x_{a,i}} = 0 \qquad \forall\, (a,i),
\end{gather*}
where $z_{a,i}$ represent inhomogeneous coordinates of the $a^{\rm th}$ point on $\mathbb{CP}^{k-1}$. More explicitly, the coordinates can be arranged in a matrix
\begin{gather*}
\begin{pmatrix}
	1 & 1 & \cdots & 1 & 1 & 0 \\
	\x_{1,1} & \x_{2,1} & \cdots & \x_{n-2,1} & \x_{n-1,1} & 0 \\
	\x_{1,2} & \x_{2,2} & \cdots & \x_{n-2,2} & \x_{n-1,2} & 0 \\
	\vdots & \vdots & \ddots & \vdots & \vdots & \vdots \\
	\x_{1,k-1} & \x_{2,k-1} & \cdots & \x_{n-2,k-1} & \x_{n-1,k-1} & 1
\end{pmatrix}\!.
\end{gather*}
Without loss of generality, here we have already used part of the redundancies inherent to this coordinates to fix that of the $n^{\rm th}$ point to infinity in a particular direction.

In Section~\ref{sec:uniqueness}, we show that on the minimal kinematics these equations possess a single solution given by the Veronese map $M_n^{k}$ introduced in \eqref{single}.

The second ingredient is what is known as the reduced determinant of the Jacobian matrix (for details, see for instance \cite[Appendix A]{Cachazo:2012pz}). The Jacobian matrix is the Hessian of the potential~${\cal S}_k$. This matrix has corank $k^2-1$ and it is usually denoted by $\Phi$. Its components,~$\Phi_{IJ}$, have composed indices $I=(a,i)$ and $J=(b,j)$ so that
\begin{gather*}
\Phi_{IJ}:=\frac{\partial^2 {\cal S}_k}{\partial \x_{a,i}\partial \x_{b,j}}.
\end{gather*}
The reduced determinant is defined by selecting a submatrix obtained from $\Phi$ by deleting $k^2-1$ rows and $k^2-1$ columns, computing its determinant and compensating with a factor which makes the object independent of the choices made. Let us denote the submatrix obtained by deleting all rows that contain labels $\{ a_1,a_2,\dots,a_{k+1}\}$, and rows containing labels $\{ b_1,b_2,\dots,b_{k+1}\}$ in~their indices by $\Phi^{a_1,a_2,\dots,a_{k+1}}_{b_1,b_2,\dots, b_{k+1}}$. Then the reduced determinant is
\begin{gather*}
{\rm det}'\Phi^{(k)} :=\frac{{\rm det}\Phi^{a_1,a_2,\dots,a_{k+1}}_{b_1,b_2,\dots, b_{k+1}}}{\V_{a_1,a_2,\dots, a_{k+1}}\V_{b_1,b_2,\dots, b_{k+1}}},
\end{gather*}
where the $\V_{a_1,a_2,\dots, a_{k+1}}$ is a generalization of a Vandermonde determinant defined by
\begin{gather*}
\V_{a_1,a_2,\dots, a_{k+1}}:= \prod_{i=1}^{k+1}|a_1,a_2,\dots, \hat{a}_{i},\dots, a_{k+1}|,
\end{gather*}
where $\hat{a}_i$ indices that the column $a_i$ has been omitted.

In Section~\ref{sec:reduced determinant}, we argue that the reduced determinant has a remarkably simple form when evaluated on the minimal kinematics and its corresponding solution to the scattering equations. The final result is simply given by
\begin{gather*}
{\rm det}'\Phi^{(k)} := \frac{1}{\big\|C_{n-1}^{(1)}\big\|^{2k-3}\big\|C_{n-1}^{(2)}\big\|^{2k-5}\cdots \big\|C_{n-1}^{(k-1)}\big\|}.
\end{gather*}

The last ingredient in the computation of the CEGM biadjoint amplitude is the so-called $k$-Parke--Taylor factor,
\begin{gather*}
{\rm PT}(1,2,\dots, n) := \frac{1}{|1,2,\dots, k||2,3,\dots, k+1|\cdots |n,1,\dots, k-1|}.
\end{gather*}
The simplest way to evaluate this object on the solution $M_n^{(k)}$ is to recall that the solution is completely symmetric in labels $\{1,2,\dots, n-1\}$. Therefore ${\rm PT}(1,2,\dots, n-1)$ is clearly cyclic in~$\p_i$ and therefore evaluates in terms of circulant graphs as
\begin{gather*}
{\rm PT}(1,2,\dots, n-1) =\frac{1}{\big\|C_{n-1}^{(1)}\big\|^{k-1}\big\|C_{n-1}^{(2)}\big\|^{k-2}\cdots \big\|C_{n-1}^{(k-1)}\big\|}.
\end{gather*}
The ratio ${\rm PT}(1,2,\dots, n-1)/{\rm PT}(1,2,\dots, n)$ is precisely the object $R_n^{(k)}$ defined in \eqref{ratioPT}.

Finally, the CHY formulation of the CEGM biadjoint amplitude is constructed as follows
\begin{gather*}
m_n^{(k)}(\mathbb{I},\mathbb{I})=\sum_{m=1}^{{\cal N}_{n,k}}\frac{1}{{\rm det}'\Phi^{(k)}}\, ({\rm PT}(1,2,\dots, n-1,n))^2\bigg|_{\x_a=\x_a^{(m)}},
\end{gather*}
where the sum runs over all ${\cal N}_{n,k}$ solutions to the scattering equations denoted $\x_a^{(m)}$.

Substituting the ingredients evaluated on the minimal kinematics and its unique solution to the scattering equations one easily finds the desired result \eqref{formi}
\begin{gather}\label{finalA}
m_n^{(k)}(\mathbb{I},\mathbb{I}) = \frac{\big(R_n^{(k)}\big)^2}{\big\|C_{n-1}^{1,2,\dots, k-1}\big\|}.
\end{gather}

Note the remarkable simplicity of this result given the fact that it has to agree with the sum over a large number of (generalized) Feynman diagrams. As~shown in Section~\ref{sec: comparison to tropGkn} for the cases avai\-lable in the literature, each (generalized) Feynman diagram coming from facets of~${\rm Trop}^+{\rm G}(k,n)$ has a non-zero contribution to the amplitude on the minimal kinematics. Moreover, individual diagrams possess a variety of poles not present in the~\eqref{finalA} as well as poles of~various orders which all cancel in the final answer to lead to only simple poles.

\section{Uniqueness of the solution}\label{sec:uniqueness}

The scattering equations are a set of polynomial equations in $(k-1)(n-(k+1))$ variables. In~general each equation obtained by differentiating ${\cal S}_n^{(k)}$ with respect to $z_{a,i}$ depends on all $n$ points on $\mathbb{CP}^{k-1}$. The key simplification that occurs for the minimal kinematics is that the range of dependence is vastly limited to neighbors of $a$. The reason is that the only non-vanishing invariants that involve $a$ are
\begin{gather*}
\s_{a-(k-1),a-(k-2),\ldots ,a},\qquad \s_{a-(k-2),a-(k-3),\ldots ,a+1},\qquad \ldots ,\qquad \s_{a,a+1,\ldots ,a+(k-1)},
\\
\s_{a-(k-2),a-(k-3),\ldots ,a,n},\qquad \s_{a-(k-3),a-(k-4),\ldots ,a+1,n},\qquad \ldots ,\qquad \s_{a,a+1,\ldots ,a-(k-2),n}.
\end{gather*}
As illustration consider $k=3$, $a=3$ and $n\geq 6$,
\begin{gather}\label{egA}
\frac{\partial {\cal S}_n^{(3)}}{\partial z_{a,i}} = \frac{\s_{123}|12|_i}{|123|} -\frac{\s_{234}|24|_i}{|234|} +\frac{\s_{345}|45|_i}{|345|}-\frac{\s_{23n}|2n|_i}{|23n|}+\frac{\s_{34n}|4n|_i}{|34n|}.
\end{gather}
Here $|ab|_i$ is the determinant of the $2\times 2$ matrix obtained by eliminating the $i^{\rm th}$ row from the $3\times 2$ matrix whose columns are the coordinates of points $a$ and $b$. The minus signs in \eqref{egA} compensate for the fact that we have chosen to write every minor in the cyclic order induced by the kinematics.

Going back to the general case, the key to easily solving the scattering equations on the minimal kinematics is a wise choice of gauge fixing. Using $\GL(k)$ and torus action redundancies in $X_{k,n}$ one can set
\begin{gather}\label{coord}
\begin{pmatrix}
	1 & 1 & \cdots & 1 & 1 & \cdots & 1 & 1 & 0 \\
	\p_1 & \p_{2} & \cdots & \p_{k} & \x_{k+1,1} & \cdots & \x_{n-2,1} & \x_{n-1,1} & 0 \\
	\p_1^2 & \p_{2}^{2} & \cdots & \p_{k}^{2} & \x_{k+1,2} & \cdots & \x_{n-2,2} & \x_{n-1,2} & 0 \\
	\vdots & \vdots & \ddots & \vdots & \vdots & \ddots & \vdots & \vdots & \vdots \\
	\p_1^{k-1} & \p_{2}^{k-1} & \cdots & \p_{k}^{k-1} & \x_{k+1,k-1} & \cdots & \x_{n-2,k-1} & \x_{n-1,k-1} & 1 \\
\end{pmatrix}\!.
\end{gather}
This is nothing but choosing a gauge in which the first $k$ columns and the last are the image of the Veronese map from the auxiliary kinematic space to $X(k,n)$. The task at hand is to show that the scattering equations localize the rest of the $\x_{a,i}$ variables to their corresponding value under the Veronese map.

The key idea is that in this gauge the scattering equations obtained by differentiating with respect to the points that are gauge fixed (after differentiation) can be turned into linear equations for some of the variables.

Let us illustrate the procedure in detail using $k=5$. The gauge fixed columns needed for~the computation are $\{ 1,2,3,4,5\}$. Differentiating the potential function with respect to~$z_{2,i}$ and~\mbox{$i\in \{1,2,3,4\}$} gives equations that can be used to solve for $z_{n-1,1}$, $z_{6,1}$, $z_{7,1}$, $z_{8,1}$. Of~par\-ti\-cu\-lar interest is
\begin{gather}\label{eq1}
z_{6,1} = \frac{z_{6,4}-s^{(1)}_{3456}z_{6,3}+s^{(2)}_{3456}z_{6,2}+s^{(4)}_{3456}}{s^{(3)}_{3456}}.
\end{gather}
Here $s^{(m)}_{3456}$ is the $m^{\rm th}$ elementary symmetric polynomial of $x_3$, $x_4$, $x_5$, $x_6$.

Repeating the same computation for the equations obtained by differentiating with res\-pect to $\{ z_{3,1},z_{3,2},z_{3,3},z_{3,4}\}$, $\{ z_{4,1},z_{4,2},z_{4,3},z_{4,4}\}$ and $\{ z_{5,1},z_{5,2},z_{5,3},z_{5,4}\}$ and solving for $z_{6,1}$ one finds
\begin{gather}
\nonumber
	z_{6,1} = \frac{z_{6,4}-s^{(1)}_{2456}z_{6,3}+s^{(2)}_{2456}z_{6,2}+s^{(4)}_{2456}}{s^{(3)}_{2456}},
\\ \nonumber
	z_{6,1} = \frac{z_{6,4}-s^{(1)}_{2356}z_{6,3}+s^{(2)}_{2356}z_{6,2}+s^{(4)}_{2356}}{s^{(3)}_{2356}},
\\
	z_{6,1} = \frac{z_{6,4}-s^{(1)}_{2346}z_{6,3}+s^{(2)}_{2346}z_{6,2}+s^{(4)}_{2346}}{s^{(3)}_{2346}},\label{eq2}
\end{gather}
respectively. Very nicely, these are four equations for the four coordinates of point $6$. The~solu\-tion to these equations is fairly easy to find by realizing that the first equation \eqref{eq1} has the same structure as
\begin{gather*}
(z-x_3)(z-x_4)(z-x_5)(z-x_6) =z^4-s_{3456}^{(1)}z^3+s_{3456}^{(2)}z^2-s_{3456}^{(3)}z+s_{3456}^{(4)}
\end{gather*}
which clearly vanishes when $z=x_6$. Likewise the other three equations in \eqref{eq2} have the structure of $(z-x_2)(z-x_4)(z-x_5)(z-x_6)$, $(z-x_2)(z-x_3)(z-x_5)(z-x_6)$ and $(z-x_2)\times(z-x_3)(z-x_4)(z-x_6)$ respectively. The only common solution is $z=x_6$ and therefore
\begin{gather*}
z_{6,1}=x_6,\qquad z_{6,2}=x_6^2,\qquad z_{6,3}=x_6^3,\qquad {\rm and}\qquad z_{6,4}=x_6^4.
\end{gather*}
Having proven that $z_{6,i} = x_{6}^i$ the whole argument can be repeated starting with the equations obtained by differentiating with respect to $\{2,3,4,5,6\}$ to find that $z_{7,i}=x_7^i$. The process ends when proving that $z_{n-1,i}=x_{n-1}^i$. This fills up the matrix \eqref{coord} to the complete Veronese map as desired.

\section{Evaluation of the reduced determinant}\label{sec:reduced determinant}

The purpose of this section is to argue that on the minimal kinematics the reduced determinant
\begin{gather*}
{\rm det}'\Phi^{(k)} :=\frac{{\rm det}\Phi^{a_1,a_2,\dots,a_{k+1}}_{b_1,b_2,\dots, b_{k+1}}}{\V_{a_1,a_2,\dots, a_{k+1}}\V_{b_1,b_2,\dots, b_{k+1}}},
\end{gather*}
evaluates to
\begin{gather}\label{expect}
{\rm det}'\Phi^{(k)} = \frac{1}{\big\|C_{n-1}^{(1)}\big\|^{2k-3}\big\|C_{n-1}^{(2)}\big\|^{2k-5}\cdots \big\|C_{n-1}^{(k-1)}\big\|}.
\end{gather}

The first observation is that the reduced determinant is permutation invariant under a replacement of labels that acts on both the kinematic invariants and the coordinates of points on $\mathbb{CP}^{k-1}$. This means that it inherits the symmetries of the minimal kinematics and that of the solutions on which it is evaluated. The minimal kinematics is cyclic in $\{x_1,x_2,\dots, x_{n-1}\}$ and so is the solution. Therefore the reduced determinant must also be cyclic. Moreover, the kinematics is also invariant under translations of $x_a$ by a constant. This means that the reduced determinant must be a function of only evaluations of circulant graph, i.e., $\big\|C_{n-1}^{(m)}\big\|$ with various values of $m$.

In order to determine the function we can consider the dependence on a single variable, say~$x_1$. The task at hand it to determine a rational function in $x_1$. Let us illustrate the procedure by studying the $k=2$ case.

Our aim is to show that ${\rm det}'\Phi^{(2)} = 1/\big\|C_{n-1}^{(1)}\big\|$. Using the freedom to eliminate three rows and columns we choose them to be $n$, 1, 2. This gives
\begin{gather*}
{\rm det}'\Phi^{(2)} = \frac{{\rm det}\,\Phi^{n,1,2}_{n,1,2}}{|n,1,2|^2}.
\end{gather*}

Let us start with poles. Clearly the only poles can be of the form $x_1-x_2$ and $x_1-x_{n-1}$. The~only entry of ${\rm det}\,\Phi^{n,1,2}_{n,1,2}$ evaluated on the minimal kinematics and its corresponding solution that depends on $x_1$ is the one in the last diagonal position, i.e. the one obtained by differentiating the potential function twice with respect to $z_{n-1}$,
\begin{gather*}
\frac{\partial^2 {\cal S}_n^{(2)}}{\partial z_{n-1}^2} = \frac{x_1-x_{n-1}}{(x_1-x_{n-1})(x_{n-1}-x_{n-2})}.
\end{gather*}
This shows that ${\rm det}\,\Phi^{n,1,2}_{n,1,2}$ has a simple pole of the form $1/(x_1-x_{n-1})$. It also shows that it must have a simple zero in $x_1-x_{2}$ since $|n,1,2|^2= (x_1-x_2)^2$ and the invariance under the change of choices made for removing columns and rows. Combining these results one has that ${\rm det}'\Phi^{(2)}$'s dependence on $x_1$ is simply $1/(x_{n-1}-x_1)(x_1-x_2)$. Using cyclicity in $n-1$ labels one concludes that ${\rm det}'\Phi^{(2)} = 1/\big\|C_{n-1}^{(1)}\big\|$. Similar analysis can be carried out for $k>2$.

We have explicitly evaluated the reduced determinant for all $(k,n)$ up to and including $(5,10)$ finding the expected result \eqref{expect}.

\section[Comparison to Trop+{\rm G}(k,n)]
{Comparison to $\boldsymbol{{\rm Trop}^+{\rm G}(k,n)}$}\label{sec: comparison to tropGkn}

CEGM biadjoint amplitudes can also be computed as a sum over generalized Feynman diagrams. These are planar arrays of Feynman diagrams which satisfy a certain compatibility condition which makes them parameterize rays in what is known as the positive Dressian ${\rm Dr}^+(k,n)$. As~announced in~\cite{Lukowski:2020dpn} and proved independently in~\cite{Arkani-Hamed:2020cig,2020arXiv200310231S}, the positive Dressian coincides with the positive Grassmannian ${\rm Trop}^+{\rm G}(k,n)$, introduced by Speyer and Williams~\cite{SWTrop}. The CEGM biadjoint amplitude can be thought of the Laplace transform of ${\rm Trop}^+{\rm G}(k,n)$. The space of kinematic invariants is the Laplace dual space to ${\rm Trop}^+{\rm G}(k,n)$. Exploring ${\rm Trop}^+{\rm G}(k,n)$ for large values of $k$ and $n$ is notoriously difficult, this is one of the motivations for introducing the minimal kinematics. Our all $k$ and $n$ result gives a peek into the structure of all positive tropical Grassmannians.

We leave the study of the implications of our results about the structure of ${\rm Trop}^+{\rm G}(k,n)$ for future work and instead concentrate here on the known cases. The very compact formulas produced from the evaluation of the CHY integral formulation imply what seems to be miraculous resummations of (generalized) Feynman diagrams.

\subsection[Resumming trees: k=2]{Resumming trees: $\boldsymbol{k=2}$}

Let us start with the $k=2$ case. Here ${\rm Trop}^+{\rm G}(2,n)$ coincides with the space of all planar cubic Feynman diagrams.\footnote{Of course, cubic Feynman diagrams can degenerate to produce higher degree vertices. This is the way one diagram can connect to another.}
Our goal is to prove that the sum over all planar Feynman diagrams evaluated on the minimal kinematics is exactly given by
\begin{gather*}
m_{n}^{(2):{\rm CHY}}(\mathbb{I},\mathbb{I}) = \frac{x_1-x_{n-1}}{(x_1-x_2)(x_2-x_3)\cdots (x_{n-2}-x_{n-1})}.
\end{gather*}
This formula was derived in~\cite{Early:2018zuw} by triangulating the type $A_{n-1}$ root cone\footnote{Also known as the dual associahedron.} and applying the homomorphism property of the continuous Laplace transform valuation on the simplicial cones in the triangulation.

Here the superscript ``CHY'' is introduced to denote the formula obtained by evaluating the CHY integral. The idea is to prove that this agrees with $m_{n}^{(2):{\rm tree}}(\mathbb{I},\mathbb{I})$, where the superscript ``tree'' is introduced to denote the sum over all planar trees.

The proof is very simple. Let us start with a computation of all planar poles on the minimal kinematics. Using momentum conservation all kinematic invariants can be expressed without using label $n$. Planar kinematic invariants are then given by
\begin{gather*}
t_a^{[r]}:=(k_a+k_{a+1}+\dots +k_{a+r})^2 = x_a-x_{a+r}.
\end{gather*}
The proof proceeds by induction and much in the same way as the BCFW recursion relations work for MHV amplitudes of gluons. This fact actually strengthens the similarities with the Parke--Taylor formula.

Let us consider the amplitude as a rational function of $x_a$ with $a\notin \{n-1,1\}$, while holding the rest generic and fixed. Let us choose $a=2$ and define $f(x_2):= m_{n}^{(2):{\rm tree}}(\mathbb{I},\mathbb{I})$.

Clearly, every planar Feynman diagram vanishes as $x_2\to \infty$ because there is at least one propagator of the form $1/(x_2-x_a)$. This is the reason for the restriction $a\notin \{n-1,1\}$. Since the planar poles are distinct from each other, $f(x_2)$ can only have simple poles. Poles are naturally separated into multi-particle and collinear.

Let us start by showing that $f(x_2)$ does not possess any multi-particle poles. Near the region where $x_2-x_a\to 0$ with $a\notin \{ 1,3\}$, only Feynman diagrams with an edge separating particles $2,3,\dots, a$ from the rest can contribute to the residue at $x_2=x_a$. The sum over such diagrams precisely turns into the product
\begin{gather*}
m^{(2):{\rm tree}}_L(2,3,\dots, a,I)\, m^{(2):{\rm tree}}_R(I, a+1,a+2,\dots, 1).
\end{gather*}
The label $I$ refers to the internal on-shell particle that appears when $x_2=x_a$. In~fact, the minimal kinematics restricted to the set $2,3,\dots, a$ can be completed into an $a$-particle physical kinematics by introducing the $I$ particle
\begin{gather*}
\begin{pmatrix}
	0 & x_2-x_3 & 0 & \cdots & 0 & x_3-x_2 \\
	x_2-x_3 & 0 & x_3-x_4 & \cdots & 0 & x_4-x_2 \\
	0 & x_3-x_4 & 0 & \cdots & 0 & x_5-x_3 \\
	\vdots & \vdots & \ddots & & \vdots & \vdots \\
	0 & 0 & 0 & \cdots & x_{a-1}-x_2 & x_2-x_{a-2} \\
	0 & 0 & 0 & \cdots & 0 & x_2-x_{a-1} \\
	x_3-x_2 & x_4-x_2 & \cdots & \cdots & x_2-x_{a-1} & 0
\end{pmatrix}\!.
\end{gather*}

This kinematics is nothing but the minimal kinematics for particles $2,3,\dots, a,I$ on the codimension one subspace given by $x_2=x_a$. Using the induction hypothesis for the points on the left amplitude, $m^{(2):{\rm tree}}_L(2,3,\dots, a,I)=m^{(2):{\rm CHY}}_L(2,3,\dots, a,I)$, one finds that the amplitude vanishes since
\begin{gather*}
m^{(2):{\rm tree}}_L(2,3,\dots, a,I) = \frac{x_2-x_{a}}{(x_2-x_3)(x_3-x_4)\cdots (x_{a-1}-x_{a})}\to 0.
\end{gather*}
This means that the residue of $f(x_2)$ on all multiparticle singularities is zero. Precisely as it is the case for MHV amplitudes of gluons.

Finally, we consider the collinear poles which are given by regions where $x_2-x_1\to 0$ or~\mbox{$x_2-x_3\to 0$}. In~each case there are exactly two Feynman diagrams that contribute and adding them up gives rise to the desired residue.

\subsection{Resumming generalized Feynman diagrams}

The Laplace transform of ${\rm Trop}^+ {\rm G}(k,n)$ can be computed using the space of planar generalized Feynman diagrams. These are nothing but arrays of metric Feynman diagrams satisfying certain compatibility conditions on their metrics. The arrays viewpoint introduced by Guevara, Umbert, Zhang and the first author allows for bootstrap approaches which have provided explicit answers for $(k,n)$ up to $(4,9)$.

Using these results we have computed and added all arrays of Feynman diagrams in the cases shown in the table on minimal kinematics and have found perfect agreement with the CHY evaluations. This is a strong consistency check on the CGUZ results.

\begin{center}\renewcommand{\arraystretch}{1.3}\setlength{\tabcolsep}{9pt}
\begin{tabular}{c|c}
\hline
$(k,n)$ & $\sharp$ of Feynman diagrams \\ \hline
$(3,6)$ & $48$ \\
$(3,7)$ & $693$ \\
$(3,8)$ & $13\, 612$ \\
$(4,8)$ & $90\, 608$ \\
\hline
\end{tabular}
\end{center}

Some comments are in order: first, we find that all diagrams give a non-zero contribution just as they did for $k=2$. This means that the CHY computation performs a surprisingly efficient resummation. The second is that while for $k=2$ all poles in the amplitude are of the form~$x_i-x_j$, the same is not true for $k>2$. This means that from the generalized Feynman diagram perspective it is not evident that if $x_1,x_2,\dots, x_{n-1}$ are kept generic the amplitude would not diverge.

We leave the important problem of deriving information regarding the structure of the full ${\rm Trop}^+ {\rm G}(k,n)$ for future work.

\section{Induced kinematics}\label{sec: induced kinematics}

Given any set of kinematic invariants for $k_1$ and $n$ one can canonically induce kinematics for $k_2$ and $n$ with $k_2 < k_1$ by summing over indices. More explicitly, if $\s_{a_1,a_2,\dots, a_{k_1}}$ satisfies momentum conservation for $(k_1,n)$ then
\begin{gather*}
\s_{a_1,a_2,\dots, a_{k_2}}:=\sum_{a_{k_2+1},a_{k_2+2},\dots, a_{k_1}=1}^n\s_{a_1,a_2,\dots, a_{k_2},a_{k_2+1},\dots, a_{k_1}}
\end{gather*}
satisfies momentum conservation for $(k_2,n)$.

Using the minimal kinematics for $k_1$ and $n$ induces different levels of minimal kinematics in lower $k$'s. We say that $k_1=k$ kinematics induces next-to-minimal kinematics (NMK) in $k_2=k-1$. More generally, $k_1=k$ induces next$^{m}$-to-minimal kinematics (N$^{m}$MK) in $k_2=k-m$.

There are two basic properties which make this an interesting generalization. The first is that the number of solutions to the scattering equations increases the farther we move from minimal kinematics, i.e., the larger the value of $m$. The second property is that for any N$^{m}$MK one can show that the Veronese embedding solution of the minimal kinematics remains a solution of the new system of equations.

We leave the study of the scattering equations on N$^{m}$MK and their solutions for future work and here we only prove the second property. The proof is very simple and in order to avoid the cluttering of notation we illustrate it using $k=4$ minimal kinematics and the N$^{m}$MK it induces.

Let us start with the equations obtained by differentiating with respect to $z_{a,3}$
\begin{gather*}
\frac{\partial {\cal S}_n^{(4)}}{\partial z_{a,3}}=\sum_{b,c,d\neq a}^n\frac{\s_{abcd}|bcd|_3}{|abcd|}.
\end{gather*}
Evaluating on the Veronese solution one finds
\begin{gather}\label{step1}
\frac{\partial {\cal S}_n^{(4)}}{\partial z_{a,3}}=\sum_{b,c,d\neq a}^n\frac{\s_{abcd}}{x_{ab}x_{ac}x_{ad}} = 0.
\end{gather}
Here we are assuming the $k=4$ minimal kinematics is being used. This is why the right hand side is zero.

Next consider the equations obtained by differentiating with respect to $z_{a,2}$ again on the Veronese solution
\begin{gather}\label{step2}
\frac{\partial {\cal S}_n^{(4)}}{\partial z_{a,2}}=\sum_{b,c,d\neq a}^n\frac{\s_{abcd}(x_b+x_c+x_d)}{x_{ab}x_{ac}x_{ad}}.
\end{gather}
The numerator can be written as $x_b+x_c+x_d=-x_{ab}-x_{ac}-x_{ad}+3x_a$. The term contai\-ning~$3x_a$ is proportional to \eqref{step1} and hence can be dropped. The other terms are all identical due to the symmetry under the exchange of labels $b$, $c $, $d$ and therefore
\begin{gather*}
\frac{\partial {\cal S}_n^{(4)}}{\partial z_{a,2}}=\sum_{b,c\neq a}^n\frac{\big(\sum_{d=1}^n\s_{abcd}\big)}{x_{ab}x_{ac}}=0.
\end{gather*}
This is nothing but the $k=3$ scattering equations for NMK evaluated on the Veronese solutions.

Finally, consider the equations obtained by differentiating with respect to $z_{a,1}$ again on the Veronese solution
\begin{gather*}
\frac{\partial {\cal S}_n^{(4)}}{\partial z_{a,1}}=\sum_{b,c,d\neq a}^n\frac{\s_{abcd}(x_bx_c+x_b x_d+x_c x_d)}{x_{ab}x_{ac}x_{ad}}.
\end{gather*}
Using the same arguments as above including \eqref{step1} and \eqref{step2} leads to the $k=2$ scattering equations on N$^{2}$MK, i.e.,
\begin{gather*}
\frac{\partial {\cal S}_n^{(4)}}{\partial z_{a,1}}=\sum_{b\neq a}^n\frac{\big(\sum_{c,d=1}^n\s_{abcd}\big)}{x_{ab}}=0.
\end{gather*}

Let us end this section with a few comments on what N$^{m}$MK looks like for $k=2$. Minimal kinematics for $k=2$ can be nicely expressed in matrix form. Here we choose to omit the $n^{\rm th}$ row and column as their entries are completely determined by momentum conservation:
\begin{gather*}
{\cal K}_n^{(2):{\rm MK}}=
\begin{pmatrix}
	0 & x_1-x_2 & 0 & \cdots & 0 & x_{n-1}-x_1 \\
	x_1-x_2 & 0 & x_2-x_3 & \cdots & 0 & 0 \\
	0 & x_2-x_3 & 0 & \cdots & 0 & 0 \\
	\vdots & \vdots & \ddots & \ddots & \ddots & \vdots \\
	0 & 0 & 0 & \cdots & 0 & x_{n-2}-x_{n-1} \\
	x_{n-1}-x_1 & 0 & 0 & \cdots & x_{n-2}-x_{n-1} & 0 \\
\end{pmatrix}\!.
\end{gather*}

Note that MK fills up the first level of entries next to the diagonal. It is easy to show that NMK corresponds to simply filling up the first two levels next to the diagonal. More explicitly, the only non-zero entries in the matrix of kinematic invariants for NMK are (with $s_{an}$ determined by momentum conservation)\footnote{That is, each row and each column should sum to zero.}
\begin{gather*}
s_{12}=x_1-x_2,\qquad s_{23}=x_{2}-x_3,\qquad \dots,\qquad s_{n-1,1}=x_{n-1}-x_1,
\\
s_{13}=x_{1}-x_{3},\qquad s_{24}=x_{2}-x_{4},\qquad\dots,\qquad s_{n-1,2}=x_{n-1}-x_2.
\end{gather*}
Likewise for N$^{m}$MK more levels are filled the more $m$ is increased.

\section[Planar-basis kinematics and multidimensional Catalan numbers]
{Planar-basis kinematics and multidimensional\\ Catalan numbers}\label{sec:PK and Ckn}

In~\cite{Cachazo:2013iea}, Cachazo, He and Yuan made the observation that the set of all possible planar kinematics invariants for $k=2$ forms a basis of the space of kinematics invariants, i.e., the set has \mbox{$\binom{n}{2}-n$} elements that can be independently chosen.\footnote{This is provided the dimension of space time is large enough. If the space-time dimension is less than the number of particles there are linear dependencies coming from the vanishing of Gram determinants which reduce the dimension of the space from $\binom{n}{2}-n$.} CHY used this observation to compute the biadjoint amplitude $m_n^{(2)}(\mathbb{I},\mathbb{I})$ on the kinematic point where all planar poles are set to unity. The motivation was two-fold. First, the computation of the amplitude as a sum over Feynman diagrams trivializes since each Feynman diagram, given by a product of planar poles, evaluates to one, and therefore $m_n^{(2)}(\mathbb{I},\mathbb{I})$ is nothing but counting the number of planar cubic trees which is known to be $C_{n-2}$, the $(n-2)^{\rm th}$ Catalan number. The second reason is that when all planar poles are set to one, all basic Mandelstam invariants, $s_{ab}$, vanish except for $2n$ of them. Note that this number is only two larger than the minimal kinematics. It turns out that the non-vanishing invariants take values
\begin{gather}%
s_{12}=s_{23}=\dots =s_{n-1,n}=s_{n,1}=1, \nonumber
\\
s_{13}=s_{24}=\dots =s_{n-1,1}=s_{n,2}=-1.\label{chyK}
\end{gather}
In fact, it is easy to see that any planar kinematic invariant evaluates to one, i.e.,
\begin{gather*}
(k_1+k_2+\cdots +k_m)^2 = s_{12}+s_{23}+\cdots + s_{m-1,m} + s_{13}+s_{24}+\cdots +s_{m-2,m} =1.
\end{gather*}

The scattering equations on this kinematics also simplify (not as dramatically as on the minimal kinematics). CHY showed that the number of solutions is $\big\lceil\frac{n-1}{2}\big\rceil$ and the equations can be mapped to a Y-system, which is a set of polynomial equations, whose solution is known. The~evaluation of the CHY formula on these solutions leads to the expected answer. At the time this was taken as evidence for the $k=2$ CHY formula which was later proven by Dolan and~Goddard in~\cite{Dolan:2013isa}.

Motivated by the $k=2$ case, in this section we discuss the generalization to higher $k$. For general $(k,n)$ the space of kinematic invariants has dimension $\binom{n}{k}-n$.

When $k>2$ is it not true that all poles appearing in the biadjoint amplitude are linearly independent as found by CEGM where $16$ planar poles were found in $(k,n)=(3,6)$
\begin{gather}\label{inv1}
\s_{123},\qquad \s_{234},\qquad \s_{345},\qquad \s_{456},\qquad \s_{561},\qquad \s_{612},
\\
\label{inv2}
t_{1234},\qquad t_{2345},\qquad t_{3456},\qquad t_{4561},\qquad t_{5612},\qquad t_{6123},
\\
\label{inv3}
R_{123456},\qquad R_{234561},\qquad {\tilde R}_{123456},\qquad {\tilde R}_{234561},
\end{gather}
and two linear relations along them,
\begin{gather}
t_{1234}+t_{3456}+t_{5612} = R_{123456} + {\tilde R}_{123456},\nonumber
\\
t_{2345}+t_{4561}+t_{6123} = R_{234561} + {\tilde R}_{234561}.\label{rela}
\end{gather}
These poles are dual to rays in ${\rm Trop}\, {\rm G}(3,6)$ and with the corresponding identifications the relations coincide with relations found by Speyer and Sturmfels in~\cite{speyer2004tropical}.

Clearly, it is impossible to set all planar kinematics invariants to unity due to \eqref{rela}. The idea is then to define a basis of planar kinematics invariants. This means that we must selected $14$ out of the $16$ possible ones. One can check that setting all $\s$ and $\t$ invariants to one and~$R$,~$\tilde R$ to~$3/2$ leads to a generic point in kinematic space and therefore the scattering equations pos\-sess~$26$ solutions.

Luckily, in~\cite{Early:2019eun} the second author introduced a planar basis of kinematic invariants with remarkable properties which come from its definition using arrangements of blades on the hyper\-simplex.

Here we show yet another remarkable property of Early's planar basis: setting all elements in the basis to unity gives rise to kinematics which naturally generalizes the CHY kinematics for $k=2$ and leads to amplitudes with a beautiful combinatorial structure.

Early's planar basis for $(3,6)$ minimally breaks the $\mathbb{Z}_2$ automorphism of the planar kinematics invariants, induced by that of ${\rm G}(3,6)$ onto itself, by selecting all elements in \eqref{inv1} and \eqref{inv2} but only the two $R$'s from \eqref{inv3}. Setting these $14$ elements to unity sets to zero all $s_{abd}$ kinematics except for $12$ of them given by
\begin{gather*}
\s_{123}=\s_{234}=\dots = \s_{561}=\s_{612}=1,
\\
\s_{613}=\s_{124}=\dots = \s_{451} = \s_{562}=-1.
\end{gather*}
This is clearly a generalization of \eqref{chyK} to $k=3$.

Let us review the construction of Early's planar basis from~\cite{Early:2019zyi,Early:2019eun}. Define linear functions $L_1,\dots, L_n$ on $\mathbb{R}^n$ by
\begin{gather*}
L_j(x) = x_{j+1} + 2x_{j+2}+\cdots (n-1)x_{j-1},
\end{gather*}
where the indices are cyclic modulo $n$.

If $J = \{j_1,\dots, j_k\}$ is a subset of $\{1,\dots, n\}$, define
\begin{gather*}
\eta_J(s) = -\frac{1}{n}\sum_{I}\s_I \min\big\{L_1(e_I-e_J),\dots, L_n(e_I-e_J)\big\},
\end{gather*}
where we abbreviate $e_J = e_{j_1}+\cdots + e_{j_k}.$

In \cite[Corollary~3.9]{Early:2019eun}, an invertible linear transformation was used to prove that the set
\begin{eqnarray}\label{eq: planar basis}
	\big\{\eta_J(s)\colon J \subset \{1,\dots, n\}, \vert J\vert = k,\ J\text{ is not a cyclic interval } \{j,j+1,\dots, j+k-1\}\big\}
\end{eqnarray}
is a basis of linear functions on the kinematic space. Moreover, if $J$ is a single cyclic interval it turns out that $\eta_J= 0$, owing to equation \eqref{cond}.
The basis in equation~\eqref{eq: planar basis} is called the \textit{planar basis.}

\begin{Definition}
	\emph{Planar basis kinematics} is the point in the kinematic space where $\eta_J=1$ for all planar basis elements.
\end{Definition}

One can derive using the inversion formula from~\cite{Early:2019eun} the direct expression in terms of Mandelstam invariants $\s_{i_1\dots i_k}$. The point is given by
\begin{gather}
\s_{12\dots k}=\s_{23\dots k+1}=\dots =\s_{n1\dots k-1}=1, \nonumber
\\
\s_{n1\dots k-2,k}=\s_{12\dots k-1,k+1}=\dots = \s_{n-1,n\dots k-3,k-1}=-1,
\label{eq: planar basis kinematics}
\end{gather}
where all other $s_J$ are set to zero.

We have been able to compute all solutions to the scattering equations for $(3,6)$, $(3,7)$ and~$(3,8)$ and evaluate the corresponding amplitude to reproduce the expected results. The~first two cases are simple exercises while the third one is more challenging. We provide the explicit solutions for $(3,8)$ in the subsection below.

Before presenting the example, let us summarize all results we have been able to obtain for the CEGM biadjoint amplitudes either by evaluating the CHY integral or by explicitly evaluating the contribution from each generalized Feynman diagram.
\begin{center}\renewcommand{\arraystretch}{1.2}\setlength{\tabcolsep}{9pt}
\begin{tabular}{c|c|c|c|c|c}
\hline
$k\setminus n$ & $4$ & $5$ & $6$ & $7$ & $8$ \\ \hline
$2$ & $2$ & $5$ & $14$ & $42$ & $132$ \\
$3$ & & $5$ & $42$ & $462$ & $6\,006$ \\
$4$ & & & $14$ & $462$ & $24\,024$ \\
$5$ & & & & $42$ & $6\, 006$ \\
$6$ & & & & & $132$ \\
\hline
\end{tabular}
\end{center}
The entry $E_{k,n}$ is the numerical value of the generalized biadjoint scalar amplitude $m^{(k)}(\mathbb{I}_n,\mathbb{I}_n)$, evaluated on the planar basis kinematics from equations \eqref{eq: planar basis kinematics}. Here $\mathbb{I}_n = (12\cdots n)$ is the standard cyclic order, as usual. In~particular, the numbers in the $k=2$ row are the Catalan numbers as explained above since they count the number of planar binary trees. Much more interesting is the fact that the remaining rows coincide with the first few values of the $k$-dimensional Catalan numbers,\footnote{Recall that the multi-dimensional Catalan number of type $(k,n)$ counts (for example) the number of rectangular standard Young tableaux of shape $k\times n$.} \cite[OEIS~A060854]{oeis}. Since the Catalan numbers had an~important meaning for~$k=2$ we expect that the same will be true for $k>2$. This is why we include data from the evaluation of the $(3,8)$ and $(4,8)$ planar arrays of Feynman diagrams in~the appendix.

\subsection[Example: CHY evaluation of (3,8)]{Example: CHY evaluation of $\boldsymbol{(3,8)}$}

Consider the convenient gauge fixing
\begin{gather}\label{allM}
\begin{pmatrix}
1 & 0 & 0 & 1 & z_{5,1} & z_{6,1} & z_{7,1} & z_{8,1}
\\[.5ex]
0 & 1 & 0 & -1 & z_{5,2} & z_{6,2} & z_{7,2} & z_{8,2}
\\[.5ex]
0 & 0 & 1 & 1 & 1 & 1 & 1 & 1
\end{pmatrix}\!.
\end{gather}
There are six solutions that can be found in this patch and one more solution at infinity. Let~us present the six explicit solutions. In~order to present them in compact form we only show the values of the $2\times 4$ matrix of variables in \eqref{allM}. As~it turns out the solutions come in pairs as they associated to three different quadratic number fields. We arrange the solutions in such pairs and label them $(1,2),(3,4),(5,6)$ indicated in the subscript of the matrices
\begin{gather*}
\begin{pmatrix}
z_{5,1} & z_{6,1} & z_{7,1} & z_{8,1}
\\
z_{5,2} & z_{6,2} & z_{7,2} & z_{8,2}
\end{pmatrix}_{1,2}\!
= \begin{pmatrix}
1\pm\frac{1}{\sqrt{2}} & 1\pm\sqrt{2} & 2\pm\sqrt{2} & 3\pm 2\sqrt{2}
\\[1ex]
\mp \sqrt{2} & -1\mp\frac{1}{\sqrt{2}} & -2 & -1\mp\sqrt{2}
\end{pmatrix}\!,
\\[1ex]
\begin{pmatrix}
z_{5,1} & z_{6,1} & z_{7,1} & z_{8,1}
\\
z_{5,2} & z_{6,2} & z_{7,2} & z_{8,2}
\end{pmatrix}_{3,4}\!
= {\setlength{\arraycolsep}{3pt}\begin{pmatrix}
2\pm \frac{\rm i}{\sqrt{2}} & 1\pm {\rm i} \sqrt{2} & \frac{1}{3} (4\pm {\rm i}\sqrt{2}) & 3
\\[1ex]
\mp \frac{1}{3} {\rm i} (\sqrt{2}\mp 4 {\rm i}) & \mp\frac{1}{6} {\rm i}(\sqrt{2}\mp 4 {\rm i}) & \pm \frac{2}{9} {\rm i}
(\sqrt{2}\pm 5 {\rm i}) & \pm \frac{1}{3} {\rm i}(\sqrt{2}\pm 5 {\rm i})
\end{pmatrix}\!,}
\\[1ex]
\begin{pmatrix}
z_{5,1} & z_{6,1} & z_{7,1} & z_{8,1}
\\
z_{5,2} & z_{6,2} & z_{7,2} & z_{8,2}
\end{pmatrix}_{5,6}\!
= \begin{pmatrix}
\frac{1}{2}\pm \frac{\rm i}{2} & 1 & 1\pm {\rm i} & 1
\\[1ex]
0 & -\frac{1}{2}\pm \frac{\rm i}{2} & 0 & \pm {\rm i}
\end{pmatrix}\!.
\end{gather*}
Computing the contribution of these six solutions to the amplitude is straightforward. Below we show the contribution of each pair of solutions in the corresponding order,
\begin{gather*}
\bigg\{\frac{47321}{8},\frac{725}{8},0\bigg\}.
\end{gather*}
Adding up the individual contributions leads to $24\, 023/4$.

Finally, the solution not captured in this patch is given by
\begin{gather}\label{spec}
\begin{pmatrix}
	1 & 1 & 1 & 1 & 1 & 1 & 1 & 1 \\
	1 & 0 & 0 & 1 & 1 & 0 & 0 & 1 \\
	0 & 1 & 0 & 1 & 0 & 1 & 0 & 1
\end{pmatrix}\!.
\end{gather}
The contribution to the amplitude is simply $1/4$. Note that the reason this is not in the patch covered by the coordinates in \eqref{allM} is that the minors $|125|$ and $|126|$ vanish.

Adding all seven contributions one finds that on the planar-basis kinematics
\begin{gather*}
m_8^{(3)}(\mathbb{I},\mathbb{I}) = 6\,006
\end{gather*}
as expected.

The attentive reader might have noticed that the structure of the seventh solution is that of two identical $3\times 4$ matrices. In~other words, points arrange themselves in pairs on the plane. The pairs are $(1,5)$, $(2,6)$, $(3,7)$ and $(4,8)$. This configuration would normally be considered singular since some minors of \eqref{spec} vanish. However, all such minors are non-planar with respect to the canonical order.

\subsection{Infinite class of solutions}

It turns out that the last solution found in the $(3,8)$ case is the first example of an infinite class of solutions. Every time $n$ is not prime, the solutions obtained for one of the factors induce solutions for $n$ by a simple replica procedure.

In order to be more explicit assume a decomposition of $n = p\, q$ with $p>k$. A solution for~$(k,p)$ induces a solution of $(k,n)$ by taking $q$ copies of the corresponding $k\times p$ matrix, $M_{k,p}$, to~create the $k\times p\, q$ matrix
\begin{gather}\label{replica}
M_{k,n}= \big( M_{k,p}\mid M_{k,p}|\cdots \mid M_{k,p}\big).
\end{gather}
The reason this is true is that the planar-basis kinematics is cyclically invariant and the scattering equations obtained by differentiating the potential function and substituting \eqref{replica} are clearly identical to those of $(k,p)$ and therefore vanish since $M_{k,p}$ is assumed to be a solution to those equations.

\section{Planar basis evaluated on the minimal kinematics}\label{sec: MK planar basis}

In this section we conjecture a formula for the evaluation of the planar basis on the minimal kinematics. For motivation and details we refer to~\cite{Early:2019zyi}.

Call a subset $J = \{j_1,\dots, j_k\}$ of $\{1,\dots, n\}$ \textit{nonfrozen} if it is not a cyclic interval $\{j,j+1,\dots, j+k-1\}$. Then \cite[Corollary~26]{Early:2019zyi} gave the following construction. Suppose $J$ has the following decomposition into $\ell$ cyclic intervals of lengths $\lambda_1,\dots, \lambda_\ell$:
\begin{gather*}
J = \lbrack j_1,j_1 + \lambda_1 -1\rbrack \cup \lbrack j_2,j_2 + \lambda_2 -1\rbrack \cup \cdots \cup \lbrack j_\ell,j_\ell + \lambda_\ell -1\rbrack.
\end{gather*}
Let $(C_1,\dots, C_\ell)$ be the interlaced complement to the intervals $I$, so that we have the concatenation
\begin{gather*}
(1,2,\dots, n) = (C_1,I_1,C_2,I_2,\dots, C_\ell,I_\ell).
\end{gather*}
Define an ordered $\ell$-tuple $(\mathbf{S}_{\mathbf{s}}) = \big((S_1)_{s_1},\dots, (S_\ell)_{s_\ell}\big)$,
where $(S_j,s_j) = \big(I_j\cup C_j , \vert I_j \vert \big)$.

The $k^\text{th}$ \textit{hypersimplex} $\Delta_{k,n}$ is the convex hull of the set of points $e_J := e_{j_1} + \cdots + e_{j_k}$ as $J$ runs over all subsets of $\{1,\dots, n\}$.
Denote by
\begin{gather*}
\big\lbrack (S_j)_{s_j},(S_{j+1})_{s_{j+1}},\dots, (S_{j-1})_{s_{j-1}}\big\rbrack
\end{gather*}
the subset\footnote{These are in fact the equations of a Schubert matroid polytope.} of $\Delta_{k,n}$ cut out by the following system of inequalities:
{\samepage\begin{gather*}
\sum_{i\in S_j}y_{i} \ge s_j,
\\
\sum_{i \in S_j\cup S_{j+1}}y_i \ge s_j+s_{j+1},
\\
\cdots\cdots\cdots\cdots\cdots\cdots\cdots\cdots
\\
\sum_{i \in S_j\cup S_{j+1}\cup \cdots \cup S_{j-2}} y_i \ge s_j+s_{j+1}+\cdots +s_{j-2}.
\end{gather*}}

In order to state our conjecture for the values of the planar basis elements $\eta_J$ on minimal kinematics, we single out two elements inside the polytope
\[
\Pi_J := \big\lbrack (S_j)_{s_j},(S_{j+1})s_{j+1},\dots, (S_{j-2})_{s_{j-2}}\big\rbrack,
\]
where the block containing the index $n$, in this case $n\in S_{j-1}$, has been eliminated.

Note that the elements $S_j\cup\cdots \cup S_{j-2}$ inherit a linear order from the cyclic order $(12\cdots n)$ by simply removing from $(12\cdots n)$ the elements in the cyclic interval $S_{j-1}$ and rotating the cyclically minimal element to the front.

If the cyclically smallest element of $S_j$ is $i_0$, say, then let $I_a$ be the interval of length $k-(s_{j-1})$ starting at $i_0$, that is
\begin{gather*}
I_a = \big\{i_0,i_0+1,\dots, k-(s_{j-1})\big\}.
\end{gather*}
Now let $I_b = J\setminus S_{j-1}$.

For instance, with $n=9$ and $k=4$, then for $J=\{2,4,7,8\}$ we have 3 cyclic intervals: 2, and 4, and 7, 8. Here $9\in \{1,2,9\}$, so
\begin{gather*}
\Pi_J = \lbrack 34_1 5678_2\rbrack.
\end{gather*}
Consequently $I_a = \{3,4,5\}$ and $I_b = \{4,7,8\}$.

With this construction we are ready to state in Conjecture \ref{conjecture} the evaluation of a planar basis element $\eta_J$ on the minimal kinematics.

\begin{Conjecture}\label{conjecture}
For any $($nonfrozen$)$ $k$-element subset $J$ of $\{1,\dots, n\}$ we have
\begin{gather*}
(\eta_J)\big\vert_{\text{MK}} = \sum_{i \in I_a}x_i - \sum_{i\in I_b} x_{i}.
\end{gather*}
Moreover, the complements $I_a\setminus I_b$ and $I_b \setminus I_a$ are non-crossing; in fact $i<j$ for all pairs $i\in I_a\setminus I_b$ and $j\in I_b \setminus I_a$. When $J$ is frozen, we have of course $\eta_J\big\vert_{MK}=0$.
\end{Conjecture}

We illustrate the conjecture in several cases, first for $k=2$.

Consider the case $(k,n) = (2,6)$ with $J=\{2,5\}$. According to the conjecture we ignore the last block of $\lbrack 345_1 126_1\rbrack$ and set $\Pi_{25} = \lbrack 345_1 \rbrack$, which is a triangle. Then $I_a = \{3,4\}$ and $I_b = \{4,5\}$, and the conjecture gives
\begin{gather*}
\eta_{25}\vert_{MK} =x_3-x_5.
\end{gather*}
In parallel, one can check directly that
\begin{gather*}
t_{345}\big\vert_{MK} = \eta_{25}\big\vert_{MK} = (s_{34}+s_{45} + s_{35})\big\vert_{MK} = x_3-x_5.
\end{gather*}

Now consider the case $(k,n) = (3,6)$ with $J_1 = \{1,4,5\}$ and $J_2 = \{2,4,6\}$. According to the conjecture,
\begin{gather*}
t_{2345}\big\vert_{MK} = \eta_{145}\big\vert_{MK} = (x_2+x_3) - (x_4+x_5).
\end{gather*}

Finally, a more nontrivial example, for $(k,n) = (4,9)$. Let us check what Conjecture \ref{conjecture} predicts for the planar pole $\eta_{1368}$. Here we set
\begin{gather*}
\Pi_J = \lbrack 23_1 456_1 78_1\rbrack,
\end{gather*}
and by explicit evaluation on the minimal kinematics equation \eqref{eq:MK} one can check that
\begin{gather*}
\eta_{1368}\big\vert_{MK} = x_2+x_4 - x_6-x_8.
\end{gather*}

\section{Discussions}

In this work we studied several kinds of special subspaces of the space of $(k,n)$ kinematic invariants with the purpose of gaining more insight into the structure of the CEGM biadjoint amplitudes as evaluated using their CHY formulation or their ${\rm Trop}^+{\rm G}(k,n)$ formulation. With~the techniques now available in the literature, both formulations are notoriously difficult to compute even for modest values of $k$ and $n$.
The minimal kinematics offers us a $n-1$ dimensional window into all values of $k$ and $n$. While the CHY formulation is relatively easy to evaluate, we do not have a similarly simple way to evaluate the arrays of~Feynman diagrams that are present in the ${\rm Trop}^+{\rm G}(k,n)$ formulation. The reason is that even on the minimal kinematics, the arrays of~Feynman diagrams have a rich structure. This is an exciting opportunity to gain some insight into ${\rm Trop}^+{\rm G}(k,n)$.

We used the general algorithms presented in the CGUZ work to evaluate the known arrays and found perfect agreement. The simplicity of the answer hints to the fact that there must be a way of understanding how the full ${\rm Trop}^+{\rm G}(k,n)$ collapses to a much simpler object.

The planar-basis kinematics is actually a single point. This very special point first explored in $k=2$ by CHY has valuable information about Feynman diagrams. The biadjoint amplitude evaluates to the number of planar binary trees. We expect that the higher $k$ generalization also has important information about generalized planar Feynman diagrams. The fact that for $k=2$ one obtains the two dimensional Catalan numbers while for $k>2$ our computations suggest that the amplitudes evaluate to the $k$-dimensional Catalan numbers\footnote{See for instance~\cite{sulanke2004generalizing}.}
\begin{gather*}
C_k(n) =\frac{1!2!\cdots (k-1)!(k(n-k))!}{(n-k)!(n-k+1)!\cdots(n-1)! },
\end{gather*}
which are known to count (for instance) certain restricted walks in a $k$-dimensional simplex, is a strong indication that they must also have a combinatorial meaning in our context. Let us simply summarize some basic facts for future work which were beyond the scope of this paper. Via an explicit bijection with rectangular standard Young tableaux, the $k$-dimensional Catalan numbers are known to enumerate the elements of the fiber over a generic point in the image of the so-called Wronski map from the Grassmannian into a projective space of the same dimension~\cite{eremenko2002degrees}; in turn the Wronski map plays a deep role in quantum integrable systems and representation theory, for instance~\cite{mukhin2009schubert,purbhoo2009jeu,speyer2012schubert}. These numbers are also related to the $h$-polynomial of the Grassmannian~\cite{braun2019hilbert}, and they count vertices of simple polytopes of dimensions $(k-1)(n-k-1)$ that generalize the usual $k=2$ (dual) associahedron~\cite{santos2017noncrossing}.

The CEGM biadjoint amplitudes, higher $k$ scattering equations, and ${\rm Trop}^+{\rm G}(k,n)$ are also deeply connected to recent studies of cluster algebras and their applications to ${\cal N}=4$ super Yang--Mills~\cite{Arkani-Hamed:2019mrd,Arkani-Hamed:2019plo,Arkani-Hamed:2019rds, Drummond:2020kqg,Drummond:2019cxm,He:2020ray,Henke:2019hve}. It would be very interesting to explore possible applications of our all $k$ and $n$ results in that context. In~particular, the all $n$ and $k=4$ answer might have information about the loop expansion in ${\cal N}=4$ super Yang--Mills.

\appendix
\section{Numerical evaluations}\label{sec:numerical}

In this appendix we collect the results of evaluating the $(3,8)$ and $(4,8)$ CEGM biadjoint amplitudes on the planar-basis kinematics defined in Section~\ref{sec:PK and Ckn}. As~mentioned in Section~\ref{sec:PK and Ckn}, when $k=2$ each planar Feynman diagram contributes $1$ to the biadjoint amplitude and therefore $m_n^{(2)}(\mathbb{I},\mathbb{I})$ counts the number of such diagrams. When $k>2$ each planar array of Feynman diagrams does not necessarily contribute $1$. The reason for this is that for $k>2$ there are more planar poles than those in the planar-basis. This has two consequences, the first is that $m_n^{(k>2)}(\mathbb{I},\mathbb{I})$ does not count the number of planar arrays of Feynman diagrams. The second is that by evaluating each array of Feynman diagrams and grouping them by their value one can hope to learn about the structure of ${\rm Trop}^+{\rm G}(k,n)$. It is the latter that motivated us to include here the data analysis for $(3,8)$ and $(4,8)$.

\subsection[(3,8) planar collections]
{$\boldsymbol{(3,8)}$ planar collections}\label{sec: planar collections 38}

There are $13\, 612$ planar collections of Feynman diagrams that contribute to $m_8^{(3)}(\mathbb{I},\mathbb{I})$. We have evaluated all of them on the planar-basis kinematics using the code provided in~\cite{Cachazo:2019xjx} and found that out of the $13\, 612$ values there are only $104$ distinct ones. Organized in ascending order they~are:
\begin{gather}\label{value}
\bigg\{ \frac{1}{144},\frac{1}{72},\frac{5}{288},\frac{1}{48},\frac{7}{288},\frac{1}{36},\frac{1}{32},
\frac{5}{144},\frac{11}{288},\frac{1}{24},\frac{7}{144},\frac{1}{18},\frac{17}{288},\frac{1}{16},
\frac{5}{72},\frac{11}{144},\frac{1}{12},\frac{13}{144},\frac{3}{32},\frac{7}{72},\nonumber
\\[1ex] \hphantom{\bigg\{}
\frac{5}{48},\frac{31}{288}, \frac{1}{9},\frac{17}{144},\frac{1}{8},\frac{5}{36},\frac{7}{48},\frac{43}{288},\frac{11}{72},\frac{5}{32}, \frac{23}{144},\frac{1}{6},\frac{25}{144},\frac{13}{72},\frac{3}{16},\frac{7}{36},\frac{29}{144}, \frac{5}{24},\frac{7}{32},\frac{2}{9},\frac{65}{288},\frac{11}{48},\nonumber
\\[1ex] \hphantom{\bigg\{}
\frac{17}{72},\frac{1}{4},\frac{37}{144},
\frac{5}{18},\frac{9}{32},\frac{7}{24},\frac{29}{96},\frac{11}{36},\frac{5}{16},\frac{1}{3},\frac{97}{288},
\frac{25}{72},\frac{17}{48},\frac{13}{36},\frac{3}{8},\frac{7}{18},\frac{13}{32},\frac{59}{144},\frac{5}{12},
\frac{31}{72},\frac{7}{16},\frac{65}{144},\nonumber
\\[1ex] \hphantom{\bigg\{}
\frac{11}{24},\frac{67}{144},\frac{15}{32},
\frac{23}{48},\frac{1}{2},\frac{13}{24},\frac{9}{16},\frac{83}{144},\frac{7}{12},\frac{29}{48},\frac{89}{144},
\frac{5}{8},\frac{23}{36},\frac{49}{72},\frac{11}{16},\frac{25}{36},\frac{101}{144},\frac{17}{24},
\frac{23}{32},\frac{3}{4},\frac{37}{48},\nonumber
\\[1ex] \hphantom{\bigg\{}
\frac{13}{16},\frac{5}{6},\frac{7}{8},\frac{11}{12},\frac{17}{18},
\frac{23}{24},1,\frac{103}{96},\frac{9}{8},\frac{169}{144},\frac{19}{16},\frac{61}{48},\frac{11}{8},
\frac{3}{2},\frac{13}{8},\frac{27}{16},\frac{7}{4},\frac{15}{8},\frac{9}{4}\bigg\} .
\end{gather}

The frequencies with which each of the values above appears is:
\begin{gather}
\{8, 24, 8, 136, 8, 16, 80, 8, 8, 328, 8, 16, 8, 480, 24, 8, 432, 8,
40, 16, 72, 8, 32, 16, 1616,\nonumber
\\ \hphantom{\{}
16, 32, 8, 32, 8, 16, 224, 24, 32, 160, 8, 8, 128, 16, 24, 8, 16, 24, 2716, 16, 8, 8, 112, 8, 8,\nonumber
\\ \hphantom{\{}
16, 72, 8, 16, 24, 8, 360, 8, 8, 8, 56, 8, 48, 16, 32, 24, 8, 16, 2632, 16, 8, 8, 16, 8, 8, 144, 8, \nonumber
\\ \hphantom{\{}
8, 40, 8, 8, 32, 8, 256, 24, 8, 16, 16, 40, 8, 8, 1972, 8, 104, 8, 8, 8, 24, 208, 8, 8, 104, 8, 12\}.\label{freq}
\end{gather}
Multiplying each value in \eqref{value} with its corresponding frequency in \eqref{freq} and totaling gives rise to the final result
\begin{gather*}
m_8^{(3)}(\mathbb{I},\mathbb{I}) = 6\,006,
\end{gather*}
which perfectly agrees with the CHY computation presented in Section~\ref{sec:PK and Ckn}.

\subsection[(4,8) planar matrices]
{$\boldsymbol{(4,8)}$ planar matrices}\label{sec: planar matrices 48}

\begin{figure}[ht!]
	\centering
\includegraphics[scale=0.97]{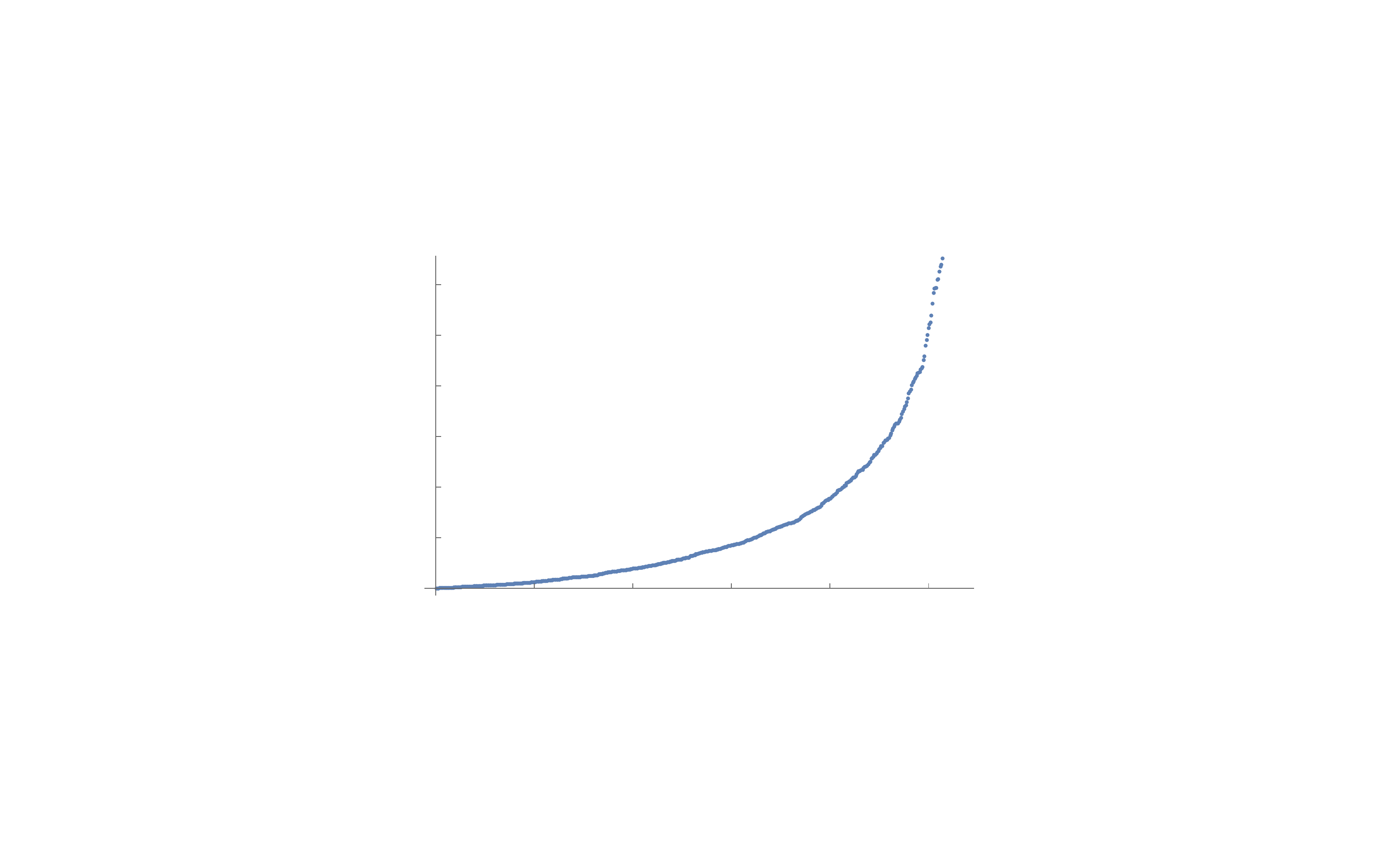}
\put(-353,192){\makebox(0,0)[lb]{\small$1.2$}}
\put(-353,160){\makebox(0,0)[lb]{\small$1.0$}}
\put(-353,129){\makebox(0,0)[lb]{\small$0.8$}}
\put(-353,97){\makebox(0,0)[lb]{\small$0.6$}}
\put(-353,67){\makebox(0,0)[lb]{\small$0.4$}}
\put(-353,35){\makebox(0,0)[lb]{\small$0.2$}}
\put(-283,-4){\makebox(0,0)[lb]{\small$100$}}
\put(-222,-4){\makebox(0,0)[lb]{\small$200$}}
\put(-161,-4){\makebox(0,0)[lb]{\small$300$}}
\put(-100,-4){\makebox(0,0)[lb]{\small$400$}}
\put(-39,-4){\makebox(0,0)[lb]{\small$500$}}
	\caption{Distinct values of planar matrices of Feynman diagrams of ${\rm Trop}^+{\rm G}(4,8)$ evaluated on the planar-basis kinematics. The $535$ values are sorted by ascending order.}
	\label{fig:Values}
\end{figure}

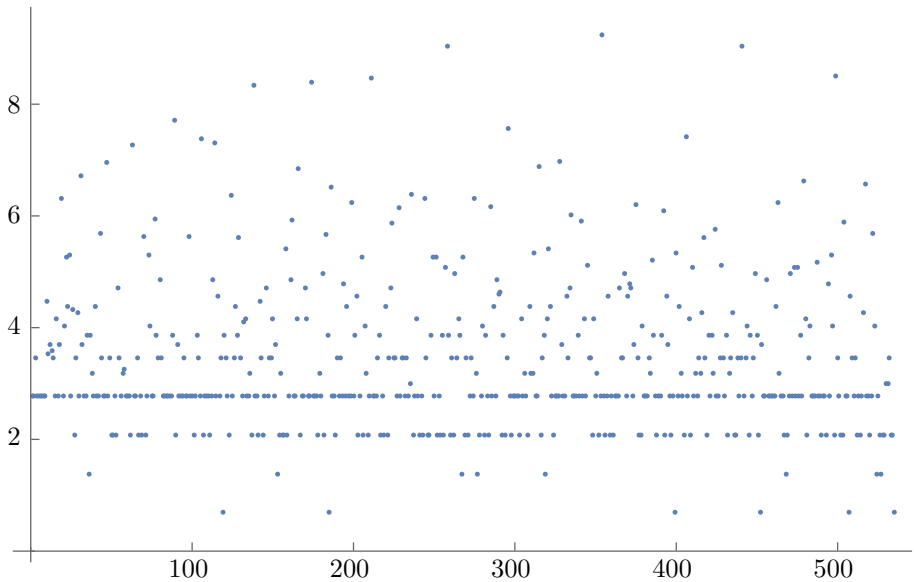
\begin{figure}[b!]
	\centering
	\includegraphics[scale=0.97]{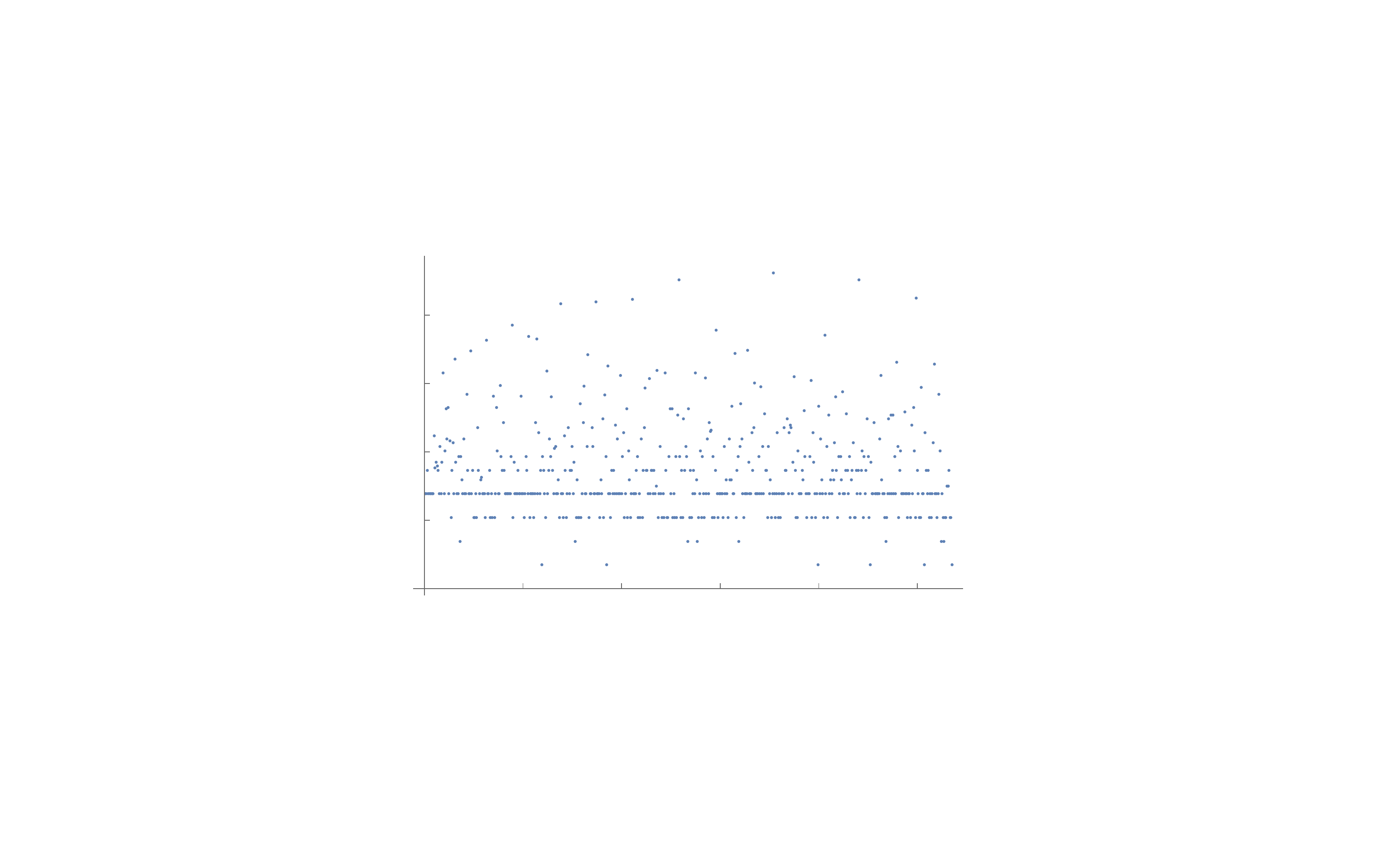}
\put(-344,172){\makebox(0,0)[lb]{\small$8$}}
\put(-344,130){\makebox(0,0)[lb]{\small$6$}}
\put(-344,88){\makebox(0,0)[lb]{\small$4$}}
\put(-344,46){\makebox(0,0)[lb]{\small$2$}}
\put(-283,-4){\makebox(0,0)[lb]{\small$100$}}
\put(-222,-4){\makebox(0,0)[lb]{\small$200$}}
\put(-161,-4){\makebox(0,0)[lb]{\small$300$}}
\put(-100,-4){\makebox(0,0)[lb]{\small$400$}}
\put(-39,-4){\makebox(0,0)[lb]{\small$500$}}
	\caption{Logarithm of frequencies of values of planar matrices of Feynman diagrams of ${\rm Trop}^+{\rm G}(4,8)$ evaluated on the planar-basis kinematics. The horizontal axis coincides with the ascending order chosen in Figure~\ref{fig:Values} to present the $535$ distinct values.}
	\label{fig:Freq}
\end{figure}

There are $90\, 608$ planar arrays of Feynman diagrams that contribute to $m_8^{(4)}(\mathbb{I},\mathbb{I})$. We have evaluated all of them on the planar-basis kinematics using the code provided in~\cite{Cachazo:2019xjx} and found that out of the $90\, 608$ values there are only $535$ distinct ones. Even though the number is small compared to the number of planar matrices, it is too large to be included here. Instead we present two plots. The first one shown in Figure~\ref{fig:Values} corresponds to the $535$ values organized in ascending order. The second one given in Figure~\ref{fig:Freq} is that of frequencies. Since frequencies vary over a large range, the plot depicts the logarithm of the corresponding frequencies.

Once again, multiplying each value shown in Figure~\ref{fig:Values} with its corresponding frequency in Figure~\ref{fig:Freq} and totaling gives rise to the final result
\begin{gather*}
m_8^{(4)}(\mathbb{I},\mathbb{I}) = 24\,024.
\end{gather*}
We have not been able to solve the scattering equations in this case so this value is a prediction for the CHY computation.

\subsection*{Acknowledgements}

We would like to thank A.~Guevara for useful discussions. This research was supported in part by a grant from the Gluskin Sheff/Onex Freeman Dyson Chair in Theoretical Physics and by Perimeter Institute. Research at Perimeter Institute is supported in part by the Government of Canada through the Department of Innovation, Science and Economic Development Canada and by the Province of Ontario through the Ministry of Colleges and Universities.

\pdfbookmark[1]{References}{ref}
\LastPageEnding

\end{document}